\documentclass[%
 aip,
 amsmath,amssymb,
 reprint,%
]{revtex4-1}

\usepackage{graphicx}% Include figure files
\usepackage{dcolumn}% Align table columns on decimal point
\usepackage{bm}% bold math
\usepackage[utf8]{inputenc}
\usepackage[T1]{fontenc}
\usepackage{mathptmx}
\usepackage{etoolbox}
\usepackage{amssymb}
\usepackage{array}

\makeatletter
\def\@email#1#2{%
 \endgroup
 \patchcmd{\titleblock@produce}
  {\frontmatter@RRAPformat}
  {\frontmatter@RRAPformat{\produce@RRAP{*#1\href{mailto:#2}{#2}}}\frontmatter@RRAPformat}
  {}{}
}%
\makeatother
\begin{document}

\preprint{AIP/123-QED}

%\title{A high-performance microwave frequency standard based on $^{113}\text{Cd}^+$
%sympathetically cooled by $^{174}\text{Yb}^+$}
\title{Precision determination of the excited-state hyperfine splitting of Cadmium ions.}
% Force line breaks with \\
\author{Y. Zheng}
\affiliation{ Department of Physics, Tsinghua University, Beijing 100084, China }
\affiliation{ State Key Laboratory of Precision Space-time Information Sensing Technology, Beijing 100084, China }% 
\author{Y. M. Yu}
\homepage{Electronic mail: ymyu@aphy.iphy.ac.cn}%Lines break automatically or can be forced with \\
\affiliation{Beijing National Laboratory for Condensed Matter Physics, Institute of Physics, Chinese Academy of Sciences, Beijing 100190,China}% 
\author{Y. T. Chen}
\affiliation{Department of Precision Instrument, Tsinghua University, Beijing 100084, China }
\affiliation{ State Key Laboratory of Precision Space-time Information Sensing Technology, Beijing 100084, China }
\author{S. N. Miao}
\affiliation{Department of Precision Instrument, Tsinghua University, Beijing 100084, China }
\affiliation{ State Key Laboratory of Precision Space-time Information Sensing Technology, Beijing 100084, China }
\author{W. X. Shi}
\affiliation{Department of Precision Instrument, Tsinghua University, Beijing 100084, China }
\affiliation{ State Key Laboratory of Precision Space-time Information Sensing Technology, Beijing 100084, China }
\author{J. W. Zhang}%
\homepage{Electronic mail: zhangjw@tsinghua.edu.cn}
\affiliation{Department of Precision Instrument, Tsinghua University, Beijing 100084, China }
\affiliation{ State Key Laboratory of Precision Space-time Information Sensing Technology, Beijing 100084, China }
\author{L. J. Wang} %\email{lwan@tsinghua.edu.cn}
\affiliation{ Department of Physics, Tsinghua University, Beijing 100084, China }
\affiliation{Department of Precision Instrument, Tsinghua University, Beijing 100084, China }
\affiliation{ State Key Laboratory of Precision Space-time Information Sensing Technology, Beijing 100084, China }


\begin{abstract}
Precision determination of the hyperfine splitting of cadmium ions is essential to study space-time variation of fundamental physical constants and isotope shifts. In this work, we present the precision frequency measurement of the excited-state $^2P_{3/2}$ hyperfine splitting of $^{111,113}\textrm{Cd}^+$ ions using the laser-induced fluorescence technique. By introducing the technology of sympathetic cooling
and setting up free-space beat detection unit based on the optical comb, the uncertainties are improved to 14.8 kHz and 10.0 kHz, respectively, two orders of magnitude higher than the reported results from the linear transformation of isotope shifts. The magnetic dipole constants $A_{P_{3/2}}$ of $^{111}\textrm{Cd}^+$ and $^{113}\textrm{Cd}^+$ are estimated to be $395~938.8(7.4)~\textrm{kHz}$ and $411~276.0(5.0)~\textrm{kHz}$, respectively. The difference between the measured and theoretical hyperfine structure constants indicates that more physical effects are required to be considered in the theoretical calculation, and provides critical data for the examination of deviation from King-plot linearity in isotope shifts.
\end{abstract}

\maketitle

Precision spectroscopy plays an important role in exploring the basic properties of atoms and studying space-time variation of fundamental physical constants. For example, the isotope shift\,(IS) provides a benchmark for testing the nuclear theory\cite{king2013isotope,silverans1988nuclear,zhao1995isotope,kong2011specific,gebert2015precision,feldker2018spectroscopy,solaro2020improved,figueroa2022precision,han2022isotope,hur2022evidence,yue2023isotope,viatkina2023calculation,imgram2019collinear,manovitz2019precision}. In particular, the deviation from King-plot linearity in isotope shifts has attracted great attention and is expected to search for new physics beyond the Standard Model\,(SM)\cite{delaunay2017probing,berengut2018probing,counts2020evidence,solaro2020improved,yerokhin2020nonlinear,rehbehn2021sensitivity,ohayon2022isotope,debierre2022testing}.
Based on the comparison between the spectra information of an element on the earth and those in the universe, the space-time variation of the fine structure constant $\alpha$ can be observed\cite{PhysRevLett.37.179,PhysRevLett.82.884,PhysRevLett.82.888}. Precision measurement of the hyperfine structure of ions, such as $\textrm{Mg}^+~$\cite{PhysRevA.96.052507,PhysRevA.107.L020803}, $\textrm{Be}^+~$\cite{PhysRevLett.101.212502}, $\textrm{Ba}^+~$\cite{becker1983precise,blatt1982precision}, $\textrm{Yb}^+~$\cite{mulholland2019laser,xin2022laser} and $\textrm{Cd}^+~$\cite{zhang2012high,miao2021precision} provides significant tests for the atom structure theory.

In recent years, the energy levels and corresponding transition lines of the cadmium ion have been widely studied in the experiment\cite{zhang2012high,zuo2019direct,miao2021precision,han2022isotope} and theory\cite{han2019roles,li2018relativistic} due to the unique advantages. Cd isotopes with nucleon numbers ranging from 100 to 130 can be produced\cite{PhysRevLett.121.102501}, and such abundant isotopes can supply valuable information on how the properties of the nucleus vary with the number of neutrons. $\textrm{Cd}^+$ ion is a good system for realizing high-performance microwave clocks\cite{zhang2012high,miao2015high,qin2022high} based on magnetic-field-insensitive transition $^2S_{1/2}\,\,|F = 0,\,\,m_F = 0⟩ \rightarrow
^2S_{1/2}\,\,|F = 1,\,\,m_F = 0⟩$ and quantum-information processing\cite{blinov2004observation}. The cadmium ion was also observed in the interstellar medium\cite{sofia1999abundance} and metal-poor stars through the Hubble Space Telescope\cite{roederer2014new}.

The ground-state hyperfine splitting\,(HFS) of trapped $^{111,113}\textrm{Cd}^+$ ions have been accurately measured using the microwave-optical double-resonance technique, to be 14 530 507 349.9(1.1) Hz and 15 199 862 855.027 99(27) Hz, respectively\cite{zhang2012high,miao2021precision}. However, the precise experimental measurement of HFS of the excited-state $^2P_{3/2}$ has rarely been reported. A preliminary value at approximately 800 MHz was available\cite{PhysRevA.53.3982}. Based on linear transformation of isotope shift of cadmium ions, the excited-state HFSs of $^{111,113}\textrm{Cd}^+$ are estimated to be 794.6(3.6) MHz and 835.5(2.9) MHz\cite{PhysRevA.106.012821}, respectively. Dixit \textit{et al}. performed the calculation of HFS in $^{113}\textrm{Cd}^+$ using the coupled-cluster theory with single, double, and partial triple excitations [CCSD(T)] method\cite{PhysRevA.77.012718}. Li \textit{et al}. extended the calculation of hyperfine structure to $^{111}\textrm{Cd}^+$ and $^{113}\textrm{Cd}^+$ employing the relativistic coupled-cluster (RCC) method\cite{PhysRevA.97.022512}. Therefore, there is a lack of direct precision measurement of excited-state HFSs of the cadmium ion.

In this paper, the HFSs of $D_2$ transition in the Cadmium ion are detected through the laser-induced fluorescence\,(LIF) technique. While measuring the transition of $5s\,\,^2S_{1/2}\,\,|F=1> \rightarrow  
 5p\,\,^2P_{3/2}\,\,|F=1,2>$, the frequency of the probe laser is scanned from red detuning to blue detuning. During this process, the trapped ions are susceptible to be heated, which will reduce the signal-to-noise ratio (SNR) and result in asymmetric transition line profiles. To avoid the situation and improve the measurement accuracy, the sympathetic cooling technique is introduced. Furthermore, the frequency of the probe laser is measured through an optical comb rather than a wavemeter to attain more accurate results. Ultimately, the magnetic dipole constants $A_{P_{3/2}}$ are given based on the measured HFSs with an accuracy two orders of magnitude higher than previous results. The precision determination of $^2P_{3/2}$-state HFSs are significant to improve the optical pump efficiency in the microwave frequency standards and quantum information processing as well as offers tests for atomic structure theory.

 The details of the linear quadrupole ion trap system used have been revealed previously\cite{zheng2023174yb+}. The ytterbium ions laser-cooled are employed as coolant to sympathetically cool cadmium ions. The relevant energy level structures of $\textrm{Yb}^+$ and $\textrm{Cd}^+$ are shown in Fig.\,1. A 369.5 nm laser ($^2S_{1/2}\,\leftrightarrow\,^2P_{1/2}$) is needed to laser-cool $^{174}\textrm{Yb}^+$ and a 935.2 nm laser is required to clear the ions spontaneously radiating to $^2D_{3/2}$ state. $^{111,113}\textrm{Cd}^+$ ions with nonzero nuclear magnetic moment $I=1/2$ have hyperfine structures. The detection is realized by a right-handed circularly polarized ($\sigma^+$) laser with a wavelength of 214.5 nm via the cycle transition $^2S_{1/2}\,\,|F=1,\,\,m_F=1>\,\leftrightarrow\,^2P_{3/2}\,\,|F=2,\,\,m_F=2>$ and the pump process is achieved by a left-handed circularly polarized ($\sigma^-$) laser via the transition $^2S_{1/2}\,\,|F=1,\,\,m_F=1>\,\rightarrow\,^2P_{3/2}\,|F=1,\,\,m_F=0>$ (written as $^2P_{3/2}\,\,|1,\,\,0>$ for simplicity in the following).

\begin{figure}[!t]    % 常规操作\begin{figure}开头说明插入图片
  \centering            % 图片放置在中间
 \includegraphics[width=0.95\linewidth]{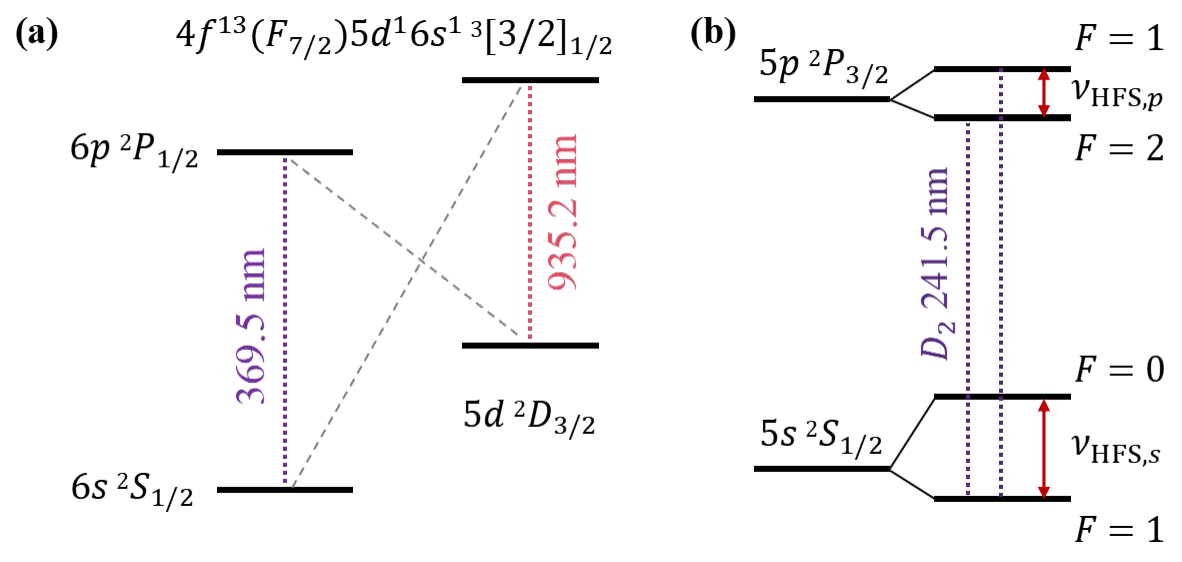}
 % \captionsetup{justification=raggedright,singlelinecheck=false}
 \caption{Relevant energy level structures of (a) ytterbium ions and (b) cadmium ions with nuclear magnetic moment $I=1/2$.}
\end{figure}

The scheme of the entire experiment setup is shown in Fig.\,2. A frequency-tripled laser (PreciLasers, Inc.) operating at 369.5 nm and an external-cavity diode laser operating at 935.2 nm are combined with a dichromatic mirror to cool and repump $\textrm{Yb}^+$ ions. 
Two 214.5 nm fourth harmonic generation lasers (TA-FHG Pro, Toptica), whose fundamental wavelengths are 858 nm, are used. One is applied to detect the fluorescence signal of $\textrm{Cd}^+$ while the other is employed as spectroscopy laser. Both lasers pass through a rochon polarizer and quarter-wave plate to obtain pure circular polarization.
In the microwave ion clock system based on $^{113}\textrm{Cd}^+$, the pump process was realized by shifting the frequency of the cooling laser using two acousto-optic modulators (AOMs)\cite{miao2021precision}. But if the frequency is scanned by AOM for around 400 MHz, there would be a large power variation, distortion of beam profile and change of propagation direction. Thus, we choose to tune the piezoelectric transducer (PZT) on the seed laser oscillator to avoid the above factors that might cause measurement uncertainties. 
The image system consists of an electron-multiplying charge-coupled device (EMCCD) and a photomultiplier tube (PMT). From the image taken by the EMCCD, the structure of sympathetically-cooled two-component ion crystal can be observed. The $\textrm{Cd}^+$ ions locate in the center and the $\textrm{Yb}^+$ ions are wrapped in the outer layer because of the greater mass as shown in Fig.\,3. 
\begin{figure}[!t]    % 常规操作\begin{figure}开头说明插入图片
  \centering            % 前面说过，图片放置在中间
 \includegraphics[width=0.95\linewidth]{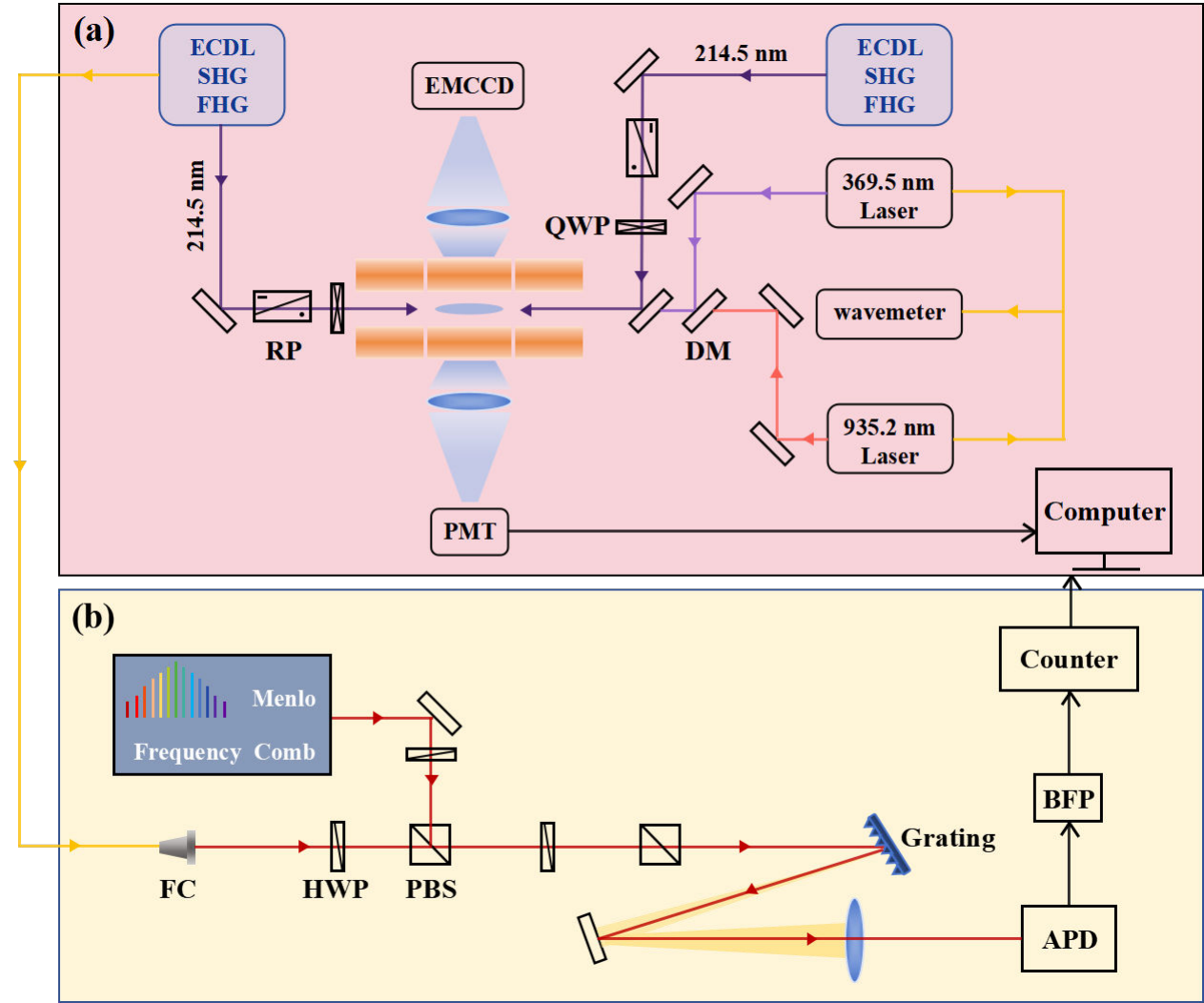}
 %\captionsetup{justification=raggedright,singlelinecheck=false}
 \caption{Experiment setup. RP, Rochon polarizer; QWP, quarter-wave plate; DM, dichromatic mirror; FC, fiber collimator; HWP, half-wave plate; PBS, polarization beam splitter; APD: Avalanche Photodetector; BPF: band-pass filter. Other elements not specified are ordinary reflection mirrors and convex lenses. The yellow lines represent the optical fibers.}
\end{figure}

\begin{figure}[!b]    % 常规操作\begin{figure}开头说明插入图片
\centering            % 前面说过，图片放置在中间
 \includegraphics[width=0.85\linewidth]{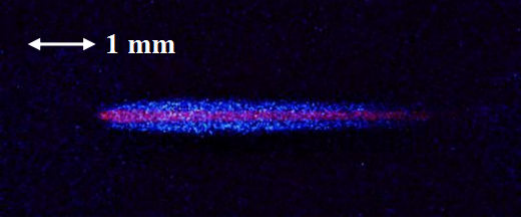}
% \captionsetup{justification=raggedright,singlelinecheck=false}
 \caption{Pseudocolor image  taken by EMCCD of $\textrm{Yb}^+-\textrm{Cd}^+$ two-component ion crystal, where the $\textrm{Yb}^+$ ions appear blue and the $\textrm{Cd}^+$ ions appear red.}
\end{figure}

\begin{figure*}[!t]    % 常规操作\begin{figure}开头说明插入图片
  \centering            % 前面说过，图片放置在中间
 \includegraphics[width=0.9\linewidth]{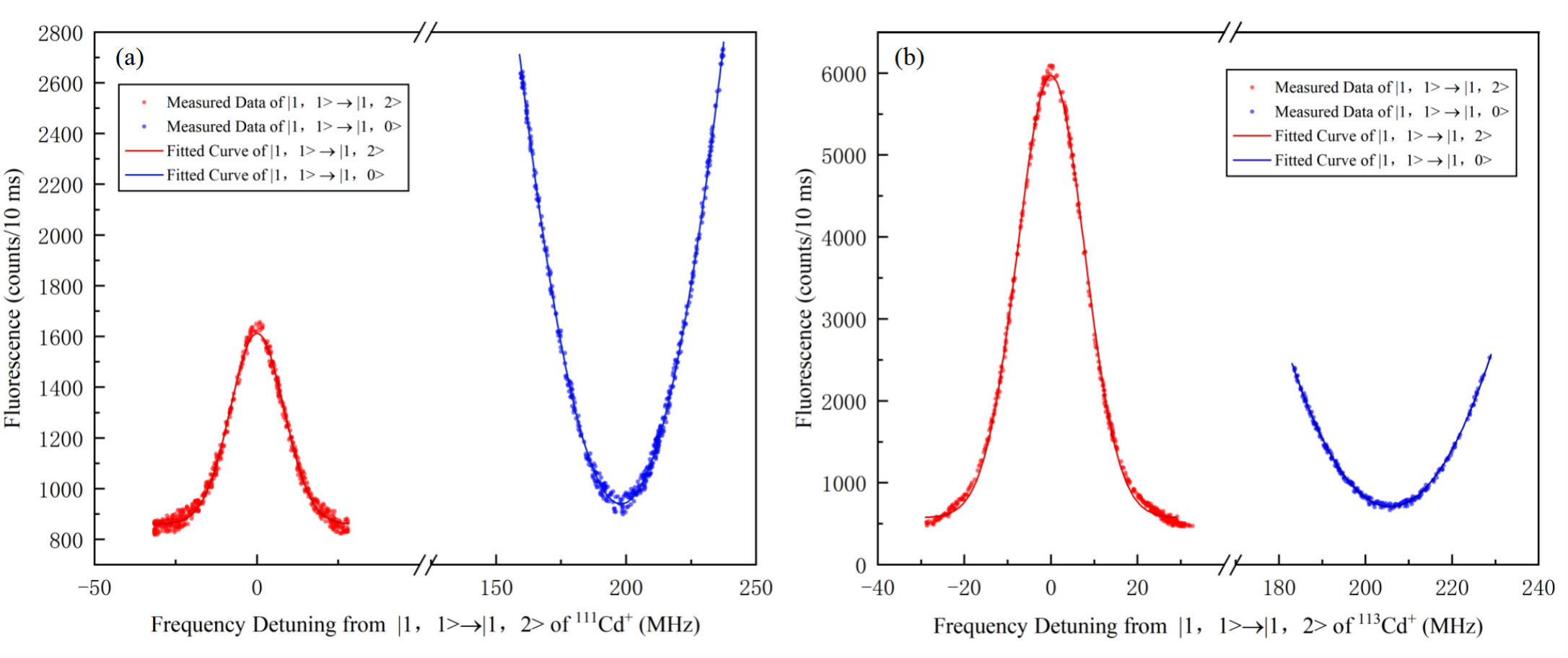}
 \caption{$D_2$ transition lines of (a) $^{111}\textrm{Cd}^+$ ion and (b) $^{113}\textrm{Cd}^+$ ion. The solid lines are the fitted profiles.}
\end{figure*}

Using a wavemeter to read out the laser frequency is a common method, but the environmental fluctuations are likely to interfere with the wavemeter and cause frequency measurement drift of even several megahertz during long-term experiment process. 
This work utilizes an optical comb to precisely measure the frequency of the 214.5 nm probe laser, greatly improving the measurement accuracy. A part of the seed laser of the fourth harmonic generation laser is coupled into the optical fiber to overlap with the comb light (Menlo Systems GmbH, FC1500-250-ULN). The Free-Space beat detection unit (BDU-FS) is set up to overlap the CW laser and comb laser, filter the broad spectrum and generate beat notes as shown in Fig.\,2(b). As the 858 nm CW laser is measured after the comb's second harmonic generation, the frequency should be calculated by formula (1):
\begin{equation}
    f_{cw}=n{\cdot}f_{rep}\,\,\pm\,\,2f_0\,\,{\pm}\,\,f_{beat},
\end{equation}
for which $f_{cw}$ is the frequency of the measured CW laser, $n$ the mode number, $f_{rep}$ the repetition rate of the comb system, $f_0$ the carrier envelop offset frequency, and $f_{beat}$ is the beat note.
The repetition rate and the offset frequency of the optical comb are phase-locked to a hydrogen maser (Microsemi, MHM 2010). The former is fixed at 250.000 061 75 MHz and the latter can be adjusted and stabilized  between 25-45 MHz as needed. The beat signal is measured by a frequency counter referenced to the same hydrogen maser. 

In order to evaluate the hyperfine splitting of the excited-state of cadmium ions, the cycling transition $^2S_{1/2}\,\,|1,\,\,1>\,\leftrightarrow\,^2P_{3/2}\,\,|2,\,\,2>$ and the pump transition $^2S_{1/2}\,\,|1,\,\,1>\,\leftrightarrow\,^2P_{3/2}\,\,|1,\,\,0>$ are measured one after the other based on the same two-component crystal to eliminate most of the systematic uncertainties. During the measurement, the intensities of the probe laser are set to be very weak at 16 $\mu \textrm{W}/\textrm{mm}^2$ for $^{113}\textrm{Cd}^+$ and 160 $\mu \textrm{W}/\textrm{mm}^2$ for $^{111}\textrm{Cd}^+$ to prevent heating or cooling the ions. 
When measuring the frequency of $^2S_{1/2}\,\,|1,\,\,1>\,\leftrightarrow\,^2P_{3/2}\,\,|2,\,\,2>$, the other 214.5 nm laser beam is blocked. The fundamental frequency of the probe laser is scanned with a range of about 100 MHz (400 MHz for FHG) around the resonance frequency. The measured line profiles of $^{111}\textrm{Cd}^+$ and $^{113}\textrm{Cd}^+$ are shown as red dots in Fig. 4(a) and Fig. 4(b), respectively, presenting an excellent symmetry. 

While detecting the frequency of pump transition, the spectroscopy laser beam is open. And its frequency is locked at the resonance frequency of transition $^2S_{1/2}\,\,|1,\,\,1>\,\leftrightarrow\,^2P_{3/2}\,\,|2,\,\,2>$. The probe laser, driving the transition between $^2S_{1/2}\,\,|1,\,\,1>$ and $^2P_{3/2}\,\,|1,\,\,0>$, can pump the ions to $^2S_{1/2}\,|F=0>$ state. If the frequency of the pump laser is far detuned from the doublet line transition, the population of $\textrm{Cd}^+$ ions remains in the cycle transition. If the laser frequency is close to the resonance frequency, the population will decay to the ground state energy level. There is no spontaneous emission, leading to a large dark signal. Every time a measurement is performed, the frequency scan is repeated multiple times to reduce the uncertainty in determining the central frequency.
These measured frequency-scanning fluorescence spectra are fitted with Gaussian profiles as different colored curves shown in Fig.\,4. The HFSs can be preliminary obtained by calculating shift of the central  frequencies of $^2S_{1/2}\,\,|1,\,\,1>\,\rightarrow\,^2P_{3/2}\,\,|2,\,\,2>$ and $^2S_{1/2}\,\,|1,\,\,1>\rightarrow\,^2P_{3/2}\,\,|1,\,\,0>$ via the following formula:
\begin{equation}
    f_{HFS}=\Delta n{\cdot}f_{rep}\pm\,2\Delta f_0\,{\pm}\,f_{beat}\,{\pm}\,f_{beat}^{'},
\end{equation}
where $\Delta n$ denotes the movement of the comb's mode number, which can be observed from a spectrum analyzer, $\Delta f_0$ is the difference in offset frequency between the two measurements, $f_{beat}$ and $f_{beat}^{'}$ are the respective beat notes of the resonance frequencies of $^2S_{1/2}\,\,|1,\,\,1>\,\rightarrow\,^2P_{3/2}\,\,|2,\,\,2>$ and $^2S_{1/2}\,\,|1,\,\,1>\,\rightarrow\,^2P_{3/2}\,\,|1,\,\,0>$ obtained from the fitted curves. The signs in the formula should be determined by slightly changing the repetition rate and the carrier envelop offset frequency before each measurement begins. For $^{111}\textrm{Cd}^+$ and $^{113}\textrm{Cd}^+$, the frequency shifts attained from BDU-FS
are 197 974.8\,(3.7) kHz and 205 643.4\,(2.5) kHz, indicating the frequency differences between the two transitions of $D_2$ line are 791 899.2(14.8) kHz and 822 573.6(10.0) kHz, respectively.

%\begin{table*}[!t]
%\caption{\label{tab:table1}%
%Uncertainties in the measured HFSs of $^2P_{3/2}$ of $^{111,113}\textrm{Cd}^+$}
%\begin{ruledtabular}
%\begin{tabular}{lccc}
 %&\textrm{ Frequency Shift\,(kHz)}&\multicolumn{2}{c}{Uncertainty\,(kHz)}\\
% Effect& &$^{111}\textrm{Cd}^+$
%&$^{113}\textrm{Cd}^+$\\ \hline
% AC Stark shift (369 nm \& 935 nm)&< 0.001&-&-\\
%%AC Stark shift (214.5 nm)&0&102.6&102.6\\
 %Zeeman shift&16.2 & 0.1&0.1 \\
 %Total type-B (systematic) uncertainty &-&102.6&102.6\\
 %Type-A (statistic) uncertainty &- &14.8&10.0\\
%Total uncertainty &-&103.6&103.1\\
%\end{tabular}
%\end{ruledtabular}
%\end{table*}

To calculate the precise excited-state HFSs of $^{111,113}\textrm{Cd}^+$, the systematic frequency shifts are carefully evaluated. The $^{111}\textrm{Cd}^+$ and $^{113}\textrm{Cd}^+$ ion have the same nucleus spin $I=1/2$, which leads to the same hyperfine structure. Therefore, the following analysis of the systematic frequency shifts applies to both $^{111}\textrm{Cd}^+$ and $^{113}\textrm{Cd}^+$.

As the measurements of $|1,\,\,1>\,\rightarrow\,|2,\,\,2>$ and $|1,\,\,1>\,\rightarrow\,|1,\,\,0>$ are performed based on the same ion crystal, the Doppler frequency shift, the Stark shift generated by the electric field of the ion trap, the blackbody radiation shift and the gravitational redshift on these two transitions are reasonable to be neglected. The uncertainty contributed by the optical frequency comb is evaluated but negligible.

Since the measurement is carried out in a weak magnetic field, the first order Zeeman frequency shift can be calculated using
\begin{equation}
       \Delta E=g_F\mu_BBM_F,
\end{equation}
 where
 \begin{equation}
  g_F=\frac{F(F+1)+J(J+1)-I(I+1)}{2F(F+1)}g^P_J,
\end{equation}
$\Delta E$ denotes the Zeeman energy, $\mu_B$ Bohr magneton, $B$ magnetic field strength, $M_F$ magnetic quantum number, $J$ the electron angular momentum, $I=1/2$ the nuclear magnetic moment of $\textrm{Cd}^+$, $F = I + J$ the total angular momentum, and $g^P_J=1.33515(43)$ the Landé g-factor of $^2P_{3/2}$ state\cite{han2019roles}.
The magnetic field strength is accurately estimated by measuring the magnetic-field-sensitive transition ($^2S_{1/2}\,\,|0,\,\,0>\,\rightarrow\,^2S_{1/2}\,\,|1,\,\,\pm1>$) of $^{113}\textrm{Cd}^+$ ions and given by:
\begin{equation}
       B=\frac{h}{\mu_B}\frac{\nu_{0-1}-\nu_{0+1}}{g^S_J+g_I},
\end{equation}
where $h$ denotes Planck constant, $g^S_J=2.002 291(4)$ and $g_I = 0.622 300 9(9) × 10^{-3}$ the Landé g-factors of $^2S_{1/2}$ state of the electron and nucleus\cite{yu2020ground,spence1972optical}, and $\nu_{0-1}$ and $\nu_{0+1}$ are the respective frequencies of magnetic-field-sensitive transitions. The frequency difference between $\nu_{0-1}$ and $\nu_{0+1}$ is 21 624(172) Hz, hence the magnitude of the magnetic field is calculated to be 771(6) nT. The corresponding frequency shift of $^2P_{3/2}\,\,|2,\,\,2>$ is 21.6(2) kHz. For the state of $^2P_{3/2}\,\,|1,\,\,0>$, there is no first order Zeeman shift. Therefore, the total uncertainty contributed by the Zeeman effect is evaluated to be 0.2 kHz.
 
As for AC-stark shift caused by cooling lasers of ytterbium ions, the quadratic Stark shift of the energy level $(F,\,\,M_F)$ takes the form \cite{angel1968hyperfine}
\begin{equation}
    \begin{split}
    \Delta \omega &= \frac{1}{2}\alpha_{sc.}(F)E^2\\
    &-\frac{1}{4}\alpha_{ten.}(F)\frac{(3M^2_F-F(F+1))}{F(2F-1)}(3F^2_z-E^2),
    \end{split}
\end{equation}
where $E$ is the isotropic part of the electric field and $E_z$ is the field in the $z$ direction, the $\alpha_{sc.}$ and $\alpha_{ten.}$ are scalar and tensor electric dipole polarizabilities, which can be estimated in terms of reduced matrix elements\cite{PhysRevA.97.022512} and transition energies \cite{nist} in the Cd$^+$ ion. The AC-stark shift is calculated to be less than 1 Hz since the lasers at the wavelength of 935 nm and 369 nm are both far-detuned for the transition frequency of cadmium ions.
The AC-stark shift due to 214.5 nm laser as resonance light for a two-level atom at rest can be evaluated based on \cite{metcalf1999laser}
\begin{eqnarray}\label{eq:Stark214}
\Delta \omega=	\bigg( \frac{\omega_{ge}}{\Gamma}\bigg)^2\frac{\delta}{1+(2\delta/\Gamma)^2},
\end{eqnarray}
where $\omega_{ge}$ and $\Gamma$ are rabi frequency and natural line width of the  $5s\,\,^2S_{1/2} \rightarrow 5p\,\,^2P_{3/2}$ transition, and $\delta$ is the detune assumed to be half of $\Gamma$. According the above formula, the shifts related to 214.5 nm laser for the transitions $^2S_{1/2}\,\,|1,\,\,1>\,\rightarrow\,^2P_{3/2}\,\,|1,\,\,0>$  and $^2S_{1/2}\,\,|1,\,\,1 >\,\rightarrow\,^2P_{3/2}\,\,|2,\,\,2>$ are reasonable to assign to zero. The uncertainty associated with the frequency instability of 214.5 nm spectroscopy laser is slight and has no influence on the final result. In summary, the total systematic frequency shift uncertainty (Type-B uncertainty) is evaluated to be 0.2 kHz.

The dominant uncertainty of the HFSs is mainly derived from the fitting uncertainty of center peaks. The statistic uncertainties (Type-A uncertainties) are determined by the square root of the sum of the squares of fitting uncertainties of the center peak positions of $|1,\,\,1>\,\rightarrow\,|1,\,\,0>$  and $|1,\,\,1 >\,\rightarrow\,|2,\,\,2>$, which are 14.8 kHz for $^{111}\textrm{Cd}^+$ and 10.0 kHz for $^{113}\textrm{Cd}^+$.
%All the systematic frequency shifts and uncertainties discussed above are listed in %Table~\ref{tab:table1}.
The HFSs of $^2P_{3/2}$-state are finally evaluated to be 791 877.6(14.8) kHz for $^{111}\textrm{Cd}^+$ and 822 552.0(10.0) kHz for $^{113}\textrm{Cd}^+$, respectively.

\begin{table*}[hbtp]
\caption{\label{tab:table2}
The hyperfine structure constants (in MHz) of $^{111,113}\textrm{Cd}^+$ ion.
}
\begin{ruledtabular}
\begin{tabular}{llccc}
 Reference&Method&$A_{P_{3/2}}^{111}$&{$A_{P_{3/2}}^{113}$}
\\ \hline
Ref.~[43]& &-&400\\
Ref.~[45]& RCC&-&406.02 \\
Ref.~[46]&  RCC &397(6)&416(6)\\
Ref.~[44]&Multicon-figuration Dirac-Hartree-Fock Method&396.0(10.5)& 415(11)\\
Ref.~[44]&Linear Transformation of Isotope Shift&397.3(1.8)&417.75(1.45)\\
 This work&LIF&395.938 8(74)&411.276 0(50)
\end{tabular}
\end{ruledtabular}
\end{table*}

For $\textrm{Cd}^+$ ion with the nuclear magnetic moment $I=1/2$, the hyperfine splitting can be described
by
\begin{equation}
    \Delta E_{P_{3/2}}=\frac{1}{2}h A_{P_{3/2}}[F(F+1)-I(I+1)-J(J+1)],
\end{equation}
where $A_{P_{3/2}}$ represents the magnetic dipole constant of the excited-state $^2P_{3/2}$. 
With the measured transition frequencies of $^2P_{3/2}\,|2, 2>$ and $^2P_{3/2}\,|1, 0>$, the hyperfine structure constants $A_{P_{3/2}}$ of $^{111,113}\textrm{Cd}^+$ are calculated to be 395 938.8(7.4) kHz and 411 276.0(5.0) kHz, respectively. 
In Table~\ref{tab:table2}, the results in this work, together with the results from other groups are listed. The uncertainty of our measurement is reduced by two orders of magnitude. There is a difference in the magnetic dipole constant between this work and Ref.~[44]. As described in that paper, we use the doublet line transition to measure the isotope shifts between $^{114}\textrm{Cd}^+$ ions and $^{111,113}\textrm{Cd}^+$ ions as well. The isotope shifts reported in Ref.~[44] were 4 649.0(1.6) MHz and 4 041.8(1.6) MHz measured using a wavemeter while our measurement results are 4 647.0516(168) MHz and 4 042.6240(132) MHz, respectively, whose uncertainties are improved by two orders of magnitude. The difference may derive from the wavemeter used in Ref.~[44] drifting more seriously than expected due to environmental influences or the deviation from King-plot linearity in isotope shifts.

In conclusion, we obtain the excited-state hyperfine splitting of the $^{111,113}\textrm{Cd}^+$ ion trapped in a linear Paul trap with LIF method. Applying the sympathetic cooling technique and setting up BDU-FS, the SNR is improved and the uncertainties of the measured HFSs are promoted by two orders of magnitude. After careful evaluation of various systematic frequency shifts, the hyperfine structure constants $A_{P_{3/2}}$ of $^{111,113}\textrm{Cd}^+$ are determined to be 395 938.8(7.4) kHz and 411 276.0(5.0) kHz, respectively. These results can be significant references for the current atomic structure calculation to take more physical effects into account. The deviation from the consequences acquired based on the linear transformation of isotope shift\cite{PhysRevA.106.012821} provides valuable test for the isotope shift theory.

\begin{acknowledgments}
This work is supported by National Key Research and Development Program of China (2021YFA1402100, 2021YFA1402104), a Project supported by the Space Application System of China Manned Space Program and National Natural Science Foundation of China (12073015).
\end{acknowledgments}
\section*{Data Availability Statement}
Data supporting the findings of this study are available upon reasonable request from the corresponding author.

\nocite{*}
\bibliography{aipsamp}% Produces the bibliography via BibTeX.

\end{document}